\documentclass[reprint,amsmath,amssymb,aps,twocolumn]{revtex4-1}
\usepackage{graphicx}
\usepackage{dcolumn}
\usepackage{bm}
\usepackage{braket}
\usepackage{hyperref}

\begin{document}
\title{A magnetic tight-binding model: surface properties of transition metals and cobalt nanoparticles}
\author{Jacques R. Eone II} 
\affiliation{%
\\ Department of Physics, University of Strasbourg,  Strasbourg, France
}%

\date{2019\\} 

\begin{abstract}
The magnetic and surface properties of some transition metals have been investigated within the tight-binding approximation, including Coulomb correlations. These surface properties are calculated after applying a charge neutrality rule that is restricted to the $d$-band. This formalism gives a charge distribution containing delocalized $sp$-states in agreement with a linear muffin-tin orbital calculation. It enables the description of local magnetism, surface energies, and work functions without recourse to the total energy. The present investigation is focused on the study of fcc cobalt, bcc iron, fcc nickel, and fcc platinum surfaces, as well as an exploration of fcc cobalt nanoparticles.
\end{abstract} 

\maketitle

\section{\label{sec:level1}Introduction}
The electronic structure of a transition metal is characterized by delocalized electrons, with a predominance of electrons in a $d$-band over those in the $sp$-band. The $d$-band is distinguished by its greater localization of electrons compared to the $sp$-band, which consists of nearly free electrons. A correlation has been observed between the two bands, and the action of the electrons occupying the $sp$-band has been shown to broaden the $d$-bandwidth and reduce the localization of the $d$-electrons \cite{cite1}.  The resultant metallic ground state can be readily described using density functional theory (DFT) \cite{cite2}. While DFT calculations are known to be efficient, their application is constrained by the number of atoms in the system under study. Therefore, semi-empirical methods, such as the tight-binding approximation, are more appropriate for studying these systems, provided that they rigorously integrate the rules governing the behavior of electrons. The tight-binding approximation is frequently restricted to the $d$-band, resulting in the neglect of the $s$- and $p$-states. This approximation yields energies that are not accurate \cite{cite1}. It is imperative to consider the impact of $sp$-band on the $d$-band when describing a transition metal. The rules that govern the electrons of a transition metal are especially important at the surface. Indeed, a clear principle of charge neutrality can be applied at the surface of a transition metal. According to some authors, the total charge is conserved at the surface of transition metals and transition metal alloys per atomic site, per orbital and per chemical species \cite{cite3,cite4,cite5}. In this study, the charge neutrality rule is restricted to the $d$-band because the $d$-electrons play a more significant role in the formation of bonds and the cohesion. By applying this charge neutrality rule to the $d$-band, the delocalized $sp$-band  generates free $sp$-states for a layer beyond the surface $(S+1)$, which represents the vacuum. The charge neutrality rule leads to a self-consistency approach useful for the calculation of surface properties like surface energies and work functions using empirical laws, as well as deducing surface magnetism. The surface magnetism is derived from the Stoner model and is obtained by shifting the nonmagnetic local density of states (LDOS). This is achieved by the assumption that the Coulomb correlations of the $d$-band are conserved at the surface. This formalism is generalized to obtain the magnetic properties of fcc cobalt nanoparticles.

\section{\label{sec:level2}Methodology}
In the tight-binding approximation, the atomic potential $H^\text{at}$ is perturbed by a weak perturbation $\Delta U(r)$ due to the interaction with  neighboring atoms. The atomic energy of the $d$-band in the atomic basis $\psi_\lambda$  can be described as shifted by a quantity $\alpha$: 
\begin{equation}
\epsilon_d =\underbrace{\int \psi_\lambda^*(r) H^\text{at}\psi_\lambda(r) d^3 r}_{\text{atomic}}+\underbrace{\int \psi_\lambda^*(r)\Delta U(r)\psi_\lambda(r) d^3 r}_{\alpha  \text{ : pertubation}}
\end{equation}

The strength of this perturbation $\Delta U(r)$ depends on the overlap between $d$ orbitals from one atom and its neighboring atoms. The coordination number is reduced at the surface, and the atomic potential is subject to different perturbations from those in the bulk due to the presence of neighboring atoms.  The impact of this novel potential at the surface can be delineated by a shift in atomic energies by a quantity $\alpha$, in accordance with a charge neutrality rule. This process is a self-consistency procedure that corrects the electronic structure at the surface prior to the study of the magnetic properties. The local magnetism is derived from the local Hubbard Hamiltonian:
\begin{equation}
H_i=- t \sum_{j,\sigma} \left( c_{i\sigma}^\dagger  c_{j\sigma} + h.c. \right)+U_d\sum_\lambda n_{\lambda\uparrow} n_{\lambda\downarrow}
\label{eq:eq1}
\end{equation}
where $\lambda$ represents a $d$ orbital, $t$ denotes the hopping integral and $U_d$ is the effective onsite Coulomb repulsion. In the context of a basis that incorporates the effects of the $s$ and $p$-states, the $d$-bandwidth is broadened and all Coulomb correlations are encompassed by $U_d$. 

\subsection{Local magnetism in the Stoner model}
The spin magnetic moment $\mu$ and the total number of electrons in the $d$-band $n_d$ can be expressed in terms of charge fluctuations \cite{cite6}: 
\begin{equation}
\mu=n_0 \left\langle n_{\uparrow} -n_{\downarrow} \right\rangle  \text{ and } n_d=n_0\left\langle n_{\uparrow} + n_{\downarrow}\right\rangle,
\end{equation}
where $n_0$=5 $d$ orbitals. The average population per spin  is approximate by: 
\[
\left\langle n_{\uparrow}  \right\rangle=\frac{1}{2n_0} (n_d-\mu)  \text{ and }\left\langle n_{\downarrow} \right\rangle=\frac{1}{2n_0} (n_d+\mu)
\]

The second term of Eq. (\ref{eq:eq1}) can be decomposed in the mean-field approximation  \cite{cite6}: 
\begin{align*}
	U_d\sum_\lambda n_{\lambda\uparrow} n_{\lambda\downarrow} &\approx U_d\sum_\lambda n_{\lambda\uparrow} \left\langle n_{\downarrow} \right\rangle + n_{\lambda\downarrow} \left\langle n_{\uparrow} \right\rangle -\left\langle n_{\uparrow} \right\rangle \left\langle n_{\downarrow} \right\rangle \\
	&=U_d\sum_{k,\sigma} n_{k\sigma} \left\langle n_{-\sigma} \right\rangle -n_0U_d \left\langle n_{\uparrow} \right\rangle \left\langle n_{\downarrow} \right\rangle \\
	&=\frac{U_d}{2n_0} \sum_{k\sigma} \left( n_d-\sigma \mu\right) c_{k\sigma}^\dagger c_{k\sigma} \\& - n_0U_d\frac{1}{4n_0^2}(n_d-\mu)(n+\mu)\\
	&=\frac{U_d}{n_0} \sum_{k\sigma} \left(\frac{n_d}{2}- \frac{\sigma }{2}\mu\right) c_{k\sigma}^\dagger c_{k\sigma} - \frac{U_d}{n_0}\left(\frac{n_d^2}{4}-\frac{\mu^2}{4}\right) 
\end{align*}

The local Hubbard Hamiltonian of  Eq. (\ref{eq:eq1})  becomes: 
\begin{equation}
H_i=\sum_{k\sigma} \left(\epsilon_k + \frac{n_dU_d}{2n_0} -\frac{\sigma }{2} \frac{U_d\mu}{n_0}\right) c_{k\sigma}^\dagger c_{k\sigma} - \frac{U_d}{n_0}\left(\frac{n_d^2}{4}-\frac{\mu^2}{4}\right) 
\label{eq:tb_etot}
\end{equation}
The band structure  $\epsilon_{k \sigma} =\epsilon_k + \frac{n_dU_d}{2n_0} -\frac{\sigma }{2}\frac{U_d\mu}{n_0} $  is then dependent on the spin  $\sigma$ and the bands are then shifted by an exchange splitting $\Delta \epsilon$. The Stoner relation is deduced as follows:
\begin{equation}
\Delta \epsilon =  \frac{U_d\mu}{n_0} = I\mu  \text{ and  } \mu=\frac{n_0}{U_d} \Delta \epsilon,
\label{eq:mu_scf}
\end{equation}
where $I$ is  the Stoner parameter. $I$ and $U$ are self-consistency parameters that are employed to obtain a correct spin magnetic moment.   The  band energy $E_b$ of a magnetic system can be derived through the summation in Eq. (\ref{eq:tb_etot}) depending on the spin:\\

\begin{equation}
E_b = \left\{
\begin{array}{ll}
\sum_{k\uparrow} \epsilon_k^\uparrow + \frac{n_dU_d}{2n_0} -\frac{1}{2} \frac{U_d\mu}{n_0} \\
\sum_{k\downarrow} \epsilon_k^\downarrow + \frac{n_dU_d}{2n_0} + \frac{1}{2} \frac{U_d\mu}{n_0} \\
\end{array}
\right. - \frac{U_d}{n_0}\left(\frac{n_d^2}{4}-\frac{\mu^2}{4}\right)
\end{equation}

The summation of the bands \textit{spin up} and \textit{spin down} containing respectively  $N_\uparrow$ et $N_\downarrow$ electrons gives:

\begin{equation}
E_b = \left\{
\begin{array}{ll}
\epsilon_{band}^\uparrow + \frac{n_dU_d}{2n_0}N_\uparrow -\frac{1}{2} \frac{U_d\mu}{n_0}N_\uparrow  \\
\epsilon_{band}^\downarrow + \frac{n_dU_d}{2n_0}N_\downarrow + \frac{1}{2} \frac{U_d\mu}{n_0}N_\downarrow \\
\end{array}
\right. - \frac{U_d}{n_0}\left(\frac{n_d^2}{4}-\frac{\mu^2}{4}\right)
\end{equation}

Or linearly:

\begin{align*}
	E_b &=  \epsilon_\text{band}^\uparrow + \epsilon_\text{band}^\downarrow + \frac{n_dU_d}{2n_0} (N_\uparrow+N_\downarrow) \\&-\frac{1}{2} \frac{U_d\mu}{n_0} (N_\uparrow-N_\downarrow) - \frac{U_d}{n_0}\left(\frac{n_d^2}{4}-\frac{\mu^2}{4}\right)\\
	&=  \epsilon_\text{band}^\uparrow + \epsilon_\text{band}^\downarrow + \frac{n_dU_d}{2n_0}n_d -\frac{1}{2} \frac{U_d\mu}{n_0}\mu - \frac{U_d}{n_0}\left(\frac{n^2}{4}-\frac{\mu^2}{4}\right) \\
	E_b &=  \epsilon_\text{band}^\uparrow + \epsilon_\text{band}^\downarrow + \frac{1}{4n_0}U_dn_d^2 -\frac{1}{4n_0}U_d\mu^2
\end{align*}

The variation in energy that occurs during the transition from a non-magnetic state to a magnetic state can be expressed as follows:
\begin{align}
	\Delta E^\text{mag.} &=  E_b^\text{mag.} - E_b^\text{nonmag.}  \\
	&= \epsilon_\text{band}^\uparrow + \epsilon_\text{band}^\downarrow - \epsilon_\text{band}^\text{nonmag.} -\frac{1}{4n_0}U_d\mu^2 \\
	\Delta E^\text{mag} &= \Delta E_{b} - \frac{1}{4n_0}U_d\mu^2
	\label{eq:e_form_mag}
\end{align}

The energy change $\Delta E^{mag}$ is negative for all ferromagnetic materials.

\subsection{Surface effects: a self-consistency treatment}
At the surface, a lower coordination number results in a decrease in bandwidth. The potential and the charge at the Fermi level are distinct from those in the bulk. The relaxation makes the distance between atoms smaller and increases the overlap between orbitals. This increases the bandwidth. Although this effect requires a total energy, it can be incorporated into a rudimentary correction. This correction involves shifting the atomic energies so that the charge and bandwidth are conserved. The electrons in the $d$-band, which are more involved in the cohesion, are used to achieve charge neutrality. Assuming that only the $d$-electrons undergo an atomic energy shift $\delta \epsilon_d$, and defining the surface Fermi level at the surface (S),  delocalized $sp$-states that lie beyond the Fermi level $(S+1)$ are obtained. This additional electronic charge, which is primarily from the $p$-band, is no longer included in the calculation of the properties of the studied surface. This procedure provides a charge distribution that is similar to the one obtained from a Linear Muffin-Tin Orbital (LMTO) calculation \cite{cite7}. Therefore, it can be concluded that the $sp$-band exerts a significant influence on the surface properties and must not be disregarded. It is evident that calculations, even when employing a total energy approach that disregards the $s$ and $p$ states, result in erroneous surface energies \cite{cite8,cite9}. The surface energy $\gamma$ is derived from an empirical law, which is defined as the difference between the band energies subsequent to the charge neutrality procedure:

\begin{equation}
\gamma = \frac{1}{3} \left[ \sum_\lambda \left(\int_{-\infty}^{E_f} E n(E,\delta\epsilon_{\lambda})dE -N_e (\lambda)\delta\epsilon_{\lambda}\right) - E_{b}^\text{bulk}\right]
\label{eq:e_surf2} 
\end{equation}

This equation is the mean value of the contribution from all the bands $\lambda = {s,p,d}$. It is posited that $\delta \epsilon_s = 0$ and  $\delta \epsilon_p = 0$. $n(E,\delta \epsilon_\lambda)$ is the shifted local density of states at the surface after the application of the charge neutrality procedure, $N_e(\lambda)$ is the number of electrons in the band $\lambda$ and $E_{b}^\text{bulk}$ is the band energy of the bulk. This expression contains a double counting  contribution $N_e (\lambda)\delta\epsilon_{i\lambda}$, which includes the energy for shifting the atomic levels of the band $\lambda$ by a quantity $\delta\epsilon_{\lambda}$. \\

The surface magnetism is derived from the Stoner model, which is applied to the LDOS. The non-magnetic LDOS is the initial point of this process, from which two LDOS spin up and spin down are created. These LDOS are then shifted by multiple values of the exchange splitting $\Delta \epsilon$ \cite{cite10}. The work function is defined at the surface as follows \cite{cite11}:

\begin{equation}
W = E_\text{vacuum} - E_F,
\label{eq:work_f} 
\end{equation}

where $E_\text{vacuum}$ denotes the energy required to extract an electron from the surface to the vacuum, without accounting for any additional kinetic energy. This vacuum energy depends on the surface properties and is derived from the mean value of the band energies subsequent to the self-consistency charge neutrality: 
\begin{equation}
E_{vacuum} = \frac{1}{3} \left[ \sum_\lambda \frac{1}{N_e (\lambda)}\left(\int_{-\infty}^{E_F} E n_{\lambda}(E,\delta\epsilon_{\lambda})dE \right) \right] - 3\gamma 
\label{eq:evacc} 
\end{equation}

In order to calculate the magnetic work function, it is necessary to add the quantity $\Delta E^{mag.}$ for the bulk and the surface. Eq (\ref{eq:evacc}) is also empirical, providing an approximate description of the work function.

\section{\label{sec:level3}Results }
\begin{figure}[!h]
	\includegraphics[width=0.3\textwidth]{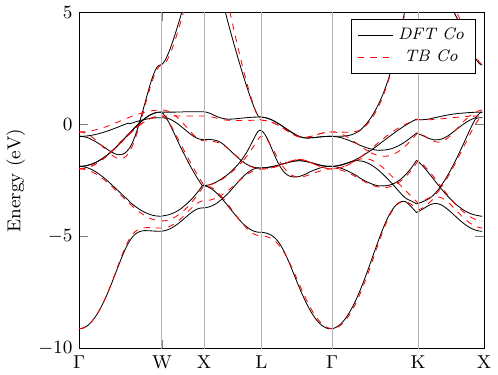}
	\caption{\label{fig:fig1} Nonmagnetic fcc Co band structures obtained using the tight-binding approximation and DFT after fitting.}
\end{figure}

\begin{figure}[!h]
	\includegraphics[width=0.3\textwidth]{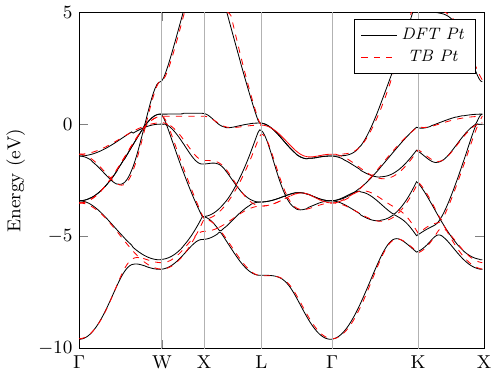}
	\caption{\label{fig:fig2} fcc Pt band structures obtained using the tight-binding approximation and DFT after fitting.}
\end{figure}

The calculations were performed using the Slater-Koster hopping parameters and atomic energies to construct the hopping integral and the tight-binding Hamiltonian.  These hopping parameters --  $ss\sigma$, $sp\sigma$, $sd\sigma$, $pp\sigma$, $pp\pi$, $pd\sigma$, $pd\pi$, $dd\sigma$, $dd\pi$, $dd\delta$ -- along with the atomic energies -- $\epsilon_s$, $\epsilon_p$ and $\epsilon_d$  -- are obtained by fitting the tight-binding band structure with the band structure obtained from a DFT calculation. The fit (Figs. \ref{fig:fig1} and \ref{fig:fig2}) and the tight-binding hamiltonian is restricted to  the first-neighbor approximation. This approximation is adequate for achieving satisfactory accuracy, particularly in the context of fcc crystal structures. Nevertheless, the hopping parameters for bcc iron are obtained from Ref. \cite{cite12}.  

\subsection{Results for bulk materials}
\begin{figure}[!h]
	\includegraphics[width=0.35\textwidth]{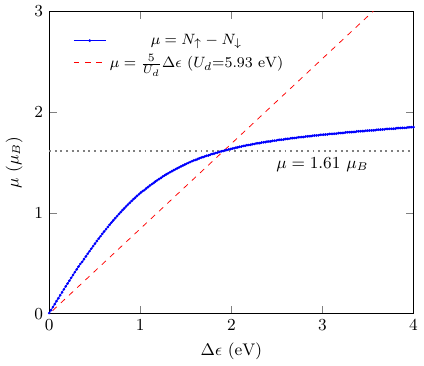}
	\caption{\label{fig:fig3} Calculation of $U_d$ for fcc Co.}
\end{figure}

The magnetic properties of ferromagnetic elements were studied using tight-binding approximation in another work \cite{cite1}. The spin magnetic moment results from the shift of nonmagnetic LDOS while keeping the electronic charge in the $d$-band $\mu = n_{d\uparrow}- n_{d\downarrow} = n_d(E-\frac{\Delta \epsilon}{2}) - n_d(E+\frac{\Delta \epsilon}{2})$. The spin magnetic moment curve is crossed by another spin magnetic moment curve from the Stoner relation in Eq. (\ref{eq:mu_scf}). This gives the value of $U_d$ in the bulk (Fig. \ref{fig:fig3}). This value should correspond to the experimental spin magnetic moment. For a $d$-band including the effects of the $sp$-band, $U_d$ is about 4.98 eV for fcc iron, 5.93 eV for fcc cobalt and 6.80 eV for fcc nickel. These values are summarized in Table \ref{table:params_corr}. The results of this study, employing new hopping parameters, exhibit slight discrepancies from those reported in a previous study \cite{cite1}. 

\begin{table}[h!]
	\caption{\label{table:params_corr} $U_d$, $\Delta \epsilon$ and $\Delta E_\text{bulk}^\text{mag}$ for ferromagnetic metals.}
	\begin{ruledtabular}
		\begin{tabular}{lccc}
		                            &  Fe     &  Co     & Ni \\ \hline 
		$U_d$ [eV]                  &   4.98   &  5.93    &  6.80  \\
		$\Delta \epsilon$ [eV]      &  2.20   & 1.91    & 0.83  \\
	    $\mu $ [$\mu_B$]            &  2.22   &  1.61   & 0.61   \\
	    $\Delta E_\text{bulk}^\text{mag}$ [eV]&  -0.28  & -0.21   & -0.03   \\
	\end{tabular}
\end{ruledtabular}
\end{table}

\subsection{Results for transition metal surfaces}
\subsubsection{Non-magnetic surface}
The application of a self-consistent procedure to the $d$-band is necessary to achieve charge neutrality at the surface (S). This results in a shift of the $d$ atomic energies by a quantity $\delta\epsilon_{d}$. This shift is dependent on the crystallographic direction (Table \ref{table:shift}). As previously stated, the surface self-consistency $d$ charge neutrality leads to the formation of free electronic charges in the $sp$-band at the Fermi level.

\begin{table}[h!]
	\caption{\label{table:shift} Energy shift of $d$ atomic levels for achieving charge neutrality in Co, Ni, Fe and Pt.}
	\begin{ruledtabular}
	\begin{tabular}{lcccc}   
		                                  &Fe       &  Co   &  Ni   &  Pt      \\ \hline 
		$\delta\epsilon_{d}(111) $ [eV]   &-        & 0.34  & 0.33  &  0.58   \\
		$\delta\epsilon_{d}(110) $ [eV]   &0.08     & -     &   -   &  -       \\
		$\delta\epsilon_{d}(100) $ [eV]   &0.30     & 0.42  & 0.43  &  0.81     \\
	\end{tabular}
\end{ruledtabular}
\end{table}

The obtained $\delta\epsilon_{d}$ value at the surface depends on the crystallographic direction. A comparison of the two crystallographic directions reveals that the $\delta\epsilon_{d}$ values for both fcc Co and fcc Ni are nearly equivalent. With regard to the distribution of charge, the Table \ref{table:charge} presents the charge at the surface (S) and the layer $(S+1)$ for fcc Co in the crystallographic directions (100) and (111). The total number of $sp$ states in the $(S+1)$ layer is consistent with an LMTO calculation for fcc Co \cite{cite7}. The same contribution to the $sp$ states at the layer $(S+1)$ is obtained for both fcc Ni and fcc Pt. However, this is insufficient to conclude that the population of this layer is constant for a crystalline structure. However,  for bcc Fe, the total contribution to $(S+1)$(100) and $(S+1)$(110) is 0.47 and 0.30 electrons respectively. A magnitude larger than in the fcc structure.  In Table \ref{table:nonmag_surf}, the nonmagnetic surface energies are calculated using  Eq. (\ref{eq:e_surf2}).

\begin{table}[h!]
	\caption{Electronic population per orbital at the surface (S) and in the layer (S+1) for a nonmagnetic fcc Co.}
	\label{table:charge}
	\begin{ruledtabular}
	\begin{tabular}{lcccc}   
	                                  &    $s$     &   $p$       &  $d$     &  Total\\ \hline 
	$N_e (100) (S) $        &  0.48    &  0.28     & 7.87     &  8.61   \\
	$N_e (100) (S+1) $     &  0.11    &  0.26     & 0.01    &  0.38   \\
	$N_e (111) (S) $        &  0.54    &  0.35     & 7.86     &  8.75    \\
	$N_e (111) (S+1) $      &  0.05    &  0.19     & 0.02     &  0.26    \\
	\end{tabular}
\end{ruledtabular}
\end{table}

\begin{table}[h!]
	\caption{\label{table:nonmag_surf}Nonmagnetic surface energies for fcc Co, fcc Ni, bcc Fe and fcc Pt.}
	\begin{ruledtabular}
		\begin{tabular}{lcccc}   
			&Fe       &  Co   &  Ni   &  Pt      \\ \hline 
			$\gamma(111) $ [eV]   &-        & 0.88  & 0.71  &  1.02   \\
			$\gamma(110) $ [eV]   &1.29     & -     &   -     &  -       \\
			$\gamma(100) $ [eV]   &0.88     & 1.19  & 0.97  &  1.43     \\
		\end{tabular}
	\end{ruledtabular}
\end{table}

The value of the surface energy $\gamma_{Fe}(100)$ is found to be greater than the experimental value of approximately 0.87 eV \cite{cite13}.  Additionally, the Pt(100) surface energy  is overestimated. A surface reconstruction occurs at this particular surface \cite{cite14}.  The calculated values are obtained using a non-reconstructed surface. The reconstruction of the Pt(100) surface  results in the formation of a hexagonal structure, which exhibits a reduced surface energy relative to the non-reconstructed surface. Notwithstanding, the Pt(111) surface energy is consistent with the experimental value of 1.03 eV \cite{cite13}, as well as a DFT calculation \cite{cite15}.

\subsubsection{Surface magnetism}
\begin{figure}[!h]
	\includegraphics[width=0.3\textwidth]{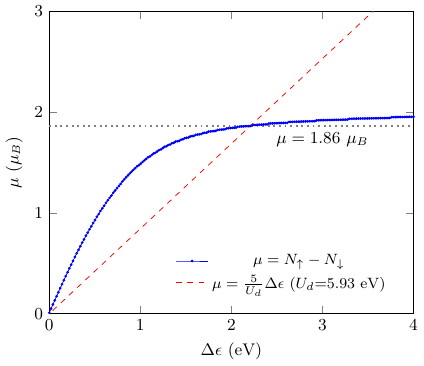}
	\caption{\label{fig:fig4}Spin magnetic moment at the surface Co(100).}
\end{figure}

It is hypothesized that charge neutrality in the $d$-band results in the conservation of the bandwidth and, consequently, the conservation of the Coulomb parameter $U_d$.  This parameter is used to determine the spin magnetic moment at the surface and the variation of energy $\Delta E^\text{mag.}$. The $U_d$ value  obtained in the bulk of fcc Co was utilized to calculate the spin magnetic moment of Co(111) and Co(100) surfaces. The resulting values of 1.77 $\mu_B$ and 1.86 $\mu_B$  (Figs. \ref{fig:fig3} and \ref{fig:fig4}) are consistent with the results reported in another calculation \cite{cite7,cite16}. Subsequent application of the identical procedure to the remaining ferromagnetic elements yielded the values presented in Table \ref{table:mag_surf}. As illustrated in Table \ref{table:mag_surf}, the spin magnetic moment at the surface is underestimated in the case of bcc Fe and overestimated in the case of fcc Ni compared to a DFT calculation. The observed discrepancies may be attributable to the charge neutrality used to describe the surface properties of these metals.\\

The magnetic surface energy can be obtained by adding to Eq. (\ref{eq:e_surf2}) the variation of the energy due to magnetism: $\delta\Delta E^{mag.}= \Delta E^{mag} (100/111/110)-\Delta E^{mag}_{bulk}$.\\

\begin{figure}[!h]
	\includegraphics[width=0.3\textwidth]{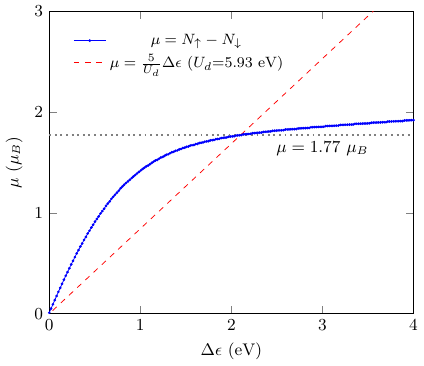}
	\caption{\label{fig:fig5}Spin magnetic moment at the surface Co(111)}
\end{figure}

 \begin{table}[h!]
 	\caption{\label{table:mag_surf}Spin magnetic moment and  $\Delta E^{mag} $ at the surface of  Co, Ni and Fe}
 	\begin{ruledtabular}
 		\begin{tabular}{lccc}   
 										            &Fe       &  Co   &  Ni   \\ \hline 
 			$\mu(111) $ [$\mu_B$]    &-        & 1.77   &  0.70 \\
 			$\mu(110) $ [$\mu_B$]   &2.54     & -     &   -       \\
 			$\mu(100) $ [$\mu_B$]   &2.65     & 1.86  &   0.83\\ \hline 
 			$\Delta E^{mag} (111) $ [eV]    &-        & -0.39  & -0.04   \\
			$\Delta E^{mag} (110) $ [eV]   &-0.52     & -     &   -       \\
			$\Delta E^{mag} (100) $ [eV]   &-0.61     & -0.45 & -0.07  \\		
 		\end{tabular}
 	\end{ruledtabular}
 \end{table}

 The magnetic surface energies are: $\gamma_{Fe}^{mag}(100)= 1.18$ eV, $\gamma_{Fe}^{mag}(110)= 0.80$ eV for bcc Fe, which are in close agreement with the experimental value of 0.89 eV \cite{cite13,cite17}. For fcc Co,  the magnetic surface energies are calculated to be $\gamma_{Co}^{mag}(100)= 0.94$ eV, $\gamma_{Co}^{mag}(111)= 0.83$ eV, which are in agreement with the experimental value of  0.87 eV  \cite{cite13,cite17}. Similarly, the magnetic surface energies for  fcc Ni are $\gamma_{Ni}^{mag}(100)= 0.94$ eV and $\gamma_{Ni}^{mag}(111)= 0.70$ eV, which are  in agreement with the experimental value of about 0.79 eV \cite{cite18} and another calculation of 0.679 eV \cite{cite16}.  \\

\begin{table*}
	\caption{Work functions in eV for Fe, Co and Ni. }
	\label{table:wf}
	\begin{ruledtabular}
	\begin{tabular}{cccccccc}
		&Fe (100)  & Fe (110)  &Co (100) & Co (111)&Ni (100) & Ni (111)& Pt(111)  \\     \hline
		$W$(Nonmag.)&  6.02    & 4.93     & 6.40    &  5.48   &  5.52  & 4.79   &   6.24     \\
		$W$(Ferro)   &  5.70     & 4.69     & 6.16    & 5.30   &  5.49   & 4.78   &     -      \\
		$W$(Expt)    &       \multicolumn{ 2}{c}{4.17  \cite{cite19} }    &        \multicolumn{ 2}{c}{5.00  \cite{cite20} }             &\multicolumn{ 2}{c}{5.15 \cite{cite20}}     &  5.65 \cite{cite20}    \\
	\end{tabular}
 	\end{ruledtabular}
\end{table*}

The calculation of work functions is achieved through the application of Eq. (\ref{eq:work_f}). As demonstrated in Table \ref{table:wf}, the calculated work functions are in close to the values obtained in another calculation \cite{cite7}. Furthermore, these values align with the experimental values. The methodology employed in the study of magnetic properties from the bulk to surfaces can be extended to the study of nanoparticles. In this study, the focus is on fcc Co nanoparticles (cuboctahedrons). However, the methodology can be extended to encompass a broader range of magnetic nanoparticles.

\subsection{Nanoparticles}
\begin{figure}[!h]
	\includegraphics[width=0.5\textwidth]{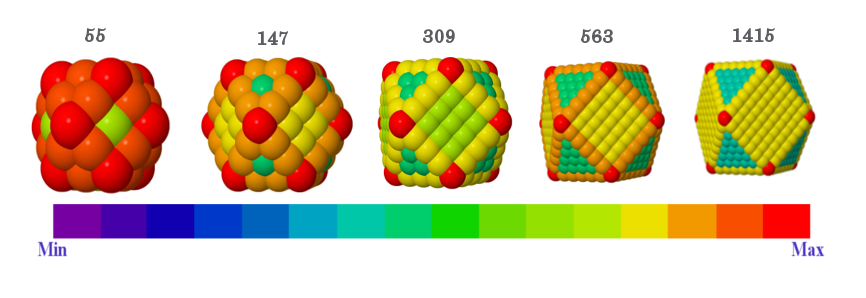}
	\caption{\label{fig:fig6}Spin magnetic moment depending on the coordination number and the size for fcc Co cuboctahedrons calculated using  $U_d$ = 5.93 eV.}
\end{figure}

\begin{figure}[!h]
 	\includegraphics[width=0.35\textwidth]{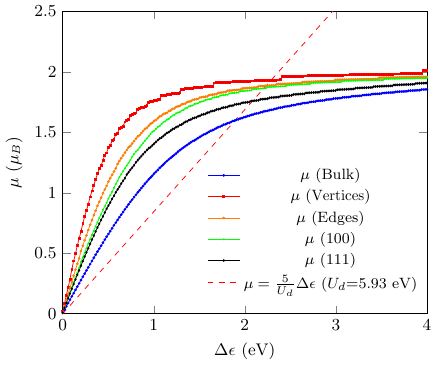}
 	\caption{\label{fig:fig7} Calculation of the spin magnetic moment for a fcc Co nanoparticle.}
 \end{figure}

The properties of a nanoparticle are determined through the assumption that all atomic sites with equivalent coordination in a first-neighbor approximation possess analogous properties. In this approximation, the nanoparticle has different classes of atomic sites.  A cuboctahedron is composed of five distinct classes of sites. The bulk (coordination number: 12), the edges (coordination number: 7), the vertices (coordination number: 5), the facet (100) (coordination number: 8), and the facet (111) (coordination number: 9). In this work, fcc Co cuboctahedrons are studied with dimensions ranging from 55 atoms to 1,415 atoms. The self-consistency surface charge neutrality procedure remains the same. A general Fermi level is established, which is predominantly defined by the bulk. The $d$ atomic energies of each class at the surface are then shifted until the charge in the $d$-band at that Fermi level is equivalent to the charge in the bulk.  The magnetic properties are calculated by shifting the nonmagnetic LDOS for each class with different values of the exchange splitting (so that five curves are defined by $\mu=N_{\uparrow}-N_{\downarrow}$). These curves are intercepted by the spin magnetic moment defined in Eq. (\ref{eq:mu_scf}) with the same Coulomb parameter $U$ = 5.93 eV (Fig. \ref{fig:fig7}). The procedure  yields a spin magnetic moment depending on the coordination depicted in Figure (\ref{fig:fig6}). A comprehensive overview  of the results is provided in Tables \ref{table:prop_nano} and \ref{table:prop_nano1}. 

 \begin{table}[h!]
 	\caption{\label{tab:table6}Spin magnetic moment, work function, surface energy for a fcc Co nanoparticle  of 1415 atoms.}
 	\label{table:prop_nano}
 	\begin{ruledtabular}
 		\begin{tabular}{cccccc}
 			                                        &Bulk      & Vertices   & Edges  &  (100) &  (111)  \\     \hline
 		 $\mu$ [$\mu_B$]               &  1.60    & 1.93       & 1.88     &  1.86    &  1.76  \\
 		            W [eV]                     &  -         & 7.99      & 6.59    &   6.09    &  5.30 \\
 	      $\gamma$  [eV]                &   -         & 1.90      & 1.31     &   1.09   &  0.81   \\
 		\end{tabular}
 	\end{ruledtabular}
 \end{table}
  \begin{table}[h!]
 	\caption{\label{tab:table6}Spin magnetic moment, work function, surface energy for a fcc Co nanoparticle of 309 atoms.}
 	\label{table:prop_nano1}
 	\begin{ruledtabular}
 		\begin{tabular}{cccccc}
 			                              &Bulk      & Vertices   & Edges  &  (100)   &  (111)  \\     \hline
 			$\mu$ [$\mu_B$]   &  1.56    & 1.89      & 1.85      &  1.81     &  1.74  \\
 			W [eV]                   &  -         & 7.64      & 6.43     &   5.89    &   5.05 \\
 			$\gamma$ [eV]      &   -         & 1.79      & 1.26     &   1.02     &  0.72  \\
 		\end{tabular}
 	\end{ruledtabular}
 \end{table}
 
Depending on their size, magnetic nanoparticles exhibit oscillatory behavior of the spin magnetic moment \cite{cite21}. This size effect  can occur in this elementary model.  This oscillation can also be observed in the variation of the work function depending on the size of the particle. Nonetheless, the surface energy  is dependent on the nanoparticle's size without a significant oscillation.

\section{\label{sec:level4}Conclusion}
In the prevailing context, the presence of nanoparticles in diverse fields has become pervasive. Consequently, the development of methodologies for the study of their properties is imperative. It is unfortunate that the most efficient approaches using \textit{ab initio} calculations are limited to approximately hundreds of atoms.  In this work, a tight-binding approach encompassing the correlations and enabling the determination of magnetic and surface properties through the application of a charge neutrality rule was introduced. The efficacy of the method is demonstrated by its capacity to produce values that are close  to those obtained from density functional theory (DFT) calculations and experimental results. This approach facilitates the calculation of the fundamental properties of magnetic and non-magnetic nanoparticles on a large scale, obviating the necessity of computing the total energy. However, the method can be extended to include the total energy, thereby enabling a more effective study of certain phenomena, such as relaxation and reconstruction. In this study, the calculation of the electronic structure was conducted in real space, thereby enabling the application of the model to non-crystalline materials or structures that exhibit defects and distortions. \\

\bibliography{article}

\end{document}